\documentclass[aps,pra,superscriptaddress,twocolumn,floats,10pt,floatfix]{revtex4-1}

\usepackage{amsmath}
\usepackage{amsfonts}
\usepackage{amssymb}
\usepackage{graphicx}
\usepackage{color}
\usepackage[colorlinks=true,citecolor=blue,linkcolor=blue]{hyperref}
\usepackage{times}
\usepackage{amstext}
\usepackage{latexsym}

\begin{document}

\title{Entangling polaritons via dynamical Casimir effect in circuit quantum electrodynamics}
\author{D. Z. Rossatto}
\affiliation{Department of Physical Chemistry, University of the Basque Country UPV/EHU, Apartado 644, E-48080 Bilbao, Spain}
\affiliation{Departamento de F\'{\i}sica, Universidade Federal de S\~{a}o Carlos, CEP 13565-905, S\~{a}o Carlos, SP, Brazil}
\author{S. Felicetti}
\author{H. Eneriz}
\affiliation{Department of Physical Chemistry, University of the Basque Country UPV/EHU, Apartado 644, E-48080 Bilbao, Spain}
\author{E. Rico}
\affiliation{Department of Physical Chemistry, University of the Basque Country UPV/EHU, Apartado 644, E-48080 Bilbao, Spain}
\affiliation{IKERBASQUE, Basque Foundation for Science, Maria Diaz de Haro 3, E-48013 Bilbao, Spain}
\author{M. Sanz}
\affiliation{Department of Physical Chemistry, University of the Basque Country UPV/EHU, Apartado 644, E-48080 Bilbao, Spain}
\author{ E. Solano}
\affiliation{Department of Physical Chemistry, University of the Basque Country UPV/EHU, Apartado 644, E-48080 Bilbao, Spain}
\affiliation{IKERBASQUE, Basque Foundation for Science, Maria Diaz de Haro 3, E-48013 Bilbao, Spain}

\begin{abstract}
We investigate theoretically how the dynamical Casimir effect can entangle quantum systems in different coupling regimes of circuit quantum electrodynamics, and show the robustness of such entanglement generation against dissipative effects, considering experimental parameters of current technology. We consider two qubit-resonator systems, which are coupled by a SQUID driven with an external magnetic field, and explore the
entire range of coupling regimes between each qubit and its resonator. In this scheme, we derive a semi-analytic explanation for the entanglement generation between both superconducting qubits when they are coupled to their resonators in the strong coupling regime. For the ultrastrong and deep strong coupling regimes, we design experimentally feasible theoretical protocols to generate maximally-entangled polaritonic states.
\end{abstract}

\maketitle

\section{Introduction}

The dynamical Casimir effect (DCE) is a quantum phenomenon that consists in the generation of detectable photons out of the modulation of the electromagnetic-field mode structure~\cite
{Moore1970,Casimir1997,Casimir1998,Dodonov2010}. Recently, the implementation of DCE in superconducting circuits~\cite{Johansson2009,Johansson2010} has been observed in two independent experiments~\cite{CMWilson2011,Hakonen2013}, by fast-modulating either the electrical boundary condition of a transmission line~\cite{CMWilson2011} or the effective speed of light in a Josephson metamaterial~\cite{Hakonen2013}. Such experiments provide a milestone in demonstrating this quantum effect, and also a feasible on-chip source of microwave entangled photon pairs.

The observation of the DCE has motivated efforts to apply it as a tool for quantum information, mainly because of the connection of the DCE to the well-known parametric amplifier from quantum optics~\cite{Casimir1998}. In particular, superconducting devices with fast-oscillating boundary conditions have been considered for entanglement generation~\cite{Felicetti2014,Aron2014,Johansson2013,Busch2014,Stassi2015}, state transfer in off-resonant couplings~\cite{Andersen2015}, quantum steering~\cite{Sabin2015}, and quantum communication~\cite{Benenti2014}. Non-Markovian dynamics~\cite{Cardenas2015}, extensions to further relativistic phenomena~\cite{FelicettiPRB2015}, implementation of optomechanical-like coupling between superconducting resonators \cite{opto1, opto2}, and ground-state attainability~\cite{Benenti2015} have also been studied in the same framework.

Aside from the fast-tuning of effective parameters, another striking feature of superconducting circuits technology is the possibility of reaching the ultrastrong coupling (USC) regime between artificial atoms and resonator modes~\cite{Bourassa2009, Niemczyk2010, Fedorov2010, Diaz2010}. When the coupling strength is comparable to the subsystems bare frequencies, the qubit and the field merge into collective bound states called polaritons~\cite{Ciuti2005}. The fast-growing interest in the USC regime is motivated by theoretical predictions of novel fundamental properties~\cite{Emary2004, Ciuti2006, Deliberato2009, Meaney2010, Ashhab2010, Ridolfo2012,deLiberato2014,Felicetti2014-3,Garziano2015}, and by potential applications in quantum computing tasks~\cite{Nataf2011, Romero2012, Kyaw2014}.

Here, in the framework of superconducting circuits, we theoretically exploit the physics underlying the DCE in order to correlate two qubit-resonator systems exploring the entire range of coupling regimes. First, we extend the numerical results of Ref.~\cite{Felicetti2014} by deriving a semi-analytic explanation for the generation of entangled states of superconducting qubits when they are strongly coupled to their resonators. Then, we generalize the analysis to include the USC regime, for which we analytically show that a maximally entangled state of the two qubit-resonator hybrid systems is feasible. To prove it, we take advantage of the strong anharmonicity and the parity protection emerging in such coupling regime. We also consider the deep strong coupling (DSC) regime~\cite{Casanova2010} in which, with the proposed scheme, a two-mode squeezed state can be produced for the two polaritonic hybrid systems. In addition, we show the robustness of such protocols of entanglement generation against dissipative effects, considering the experimental parameters of the current technology for strong coupling (SC) and USC regimes. Then, we show that this protocol, which can be implemented in an on-chip architecture \cite{Felicetti2014} without requiring external squeezed signals \cite{Kraus2004}, is a compelling tool for quantum networks even when each qubit-resonator block of a network internally works in either ultrastrong or deep strong coupling regimes.

The paper is organized as follows. In Sec. II, we introduce a model describing the two qubit-resonator systems under DCE action. Section III outlines the entanglement protocols for the entire range of coupling regimes, involving the SC regime in Sec. III A, the USC regime in Sec. III B, and the DSC regime in Sec. III C. Finally, Sec. IV covers the summary of the results.

\section{Theoretical Model}

\begin{figure}[t]
\includegraphics[width=0.45\textwidth]{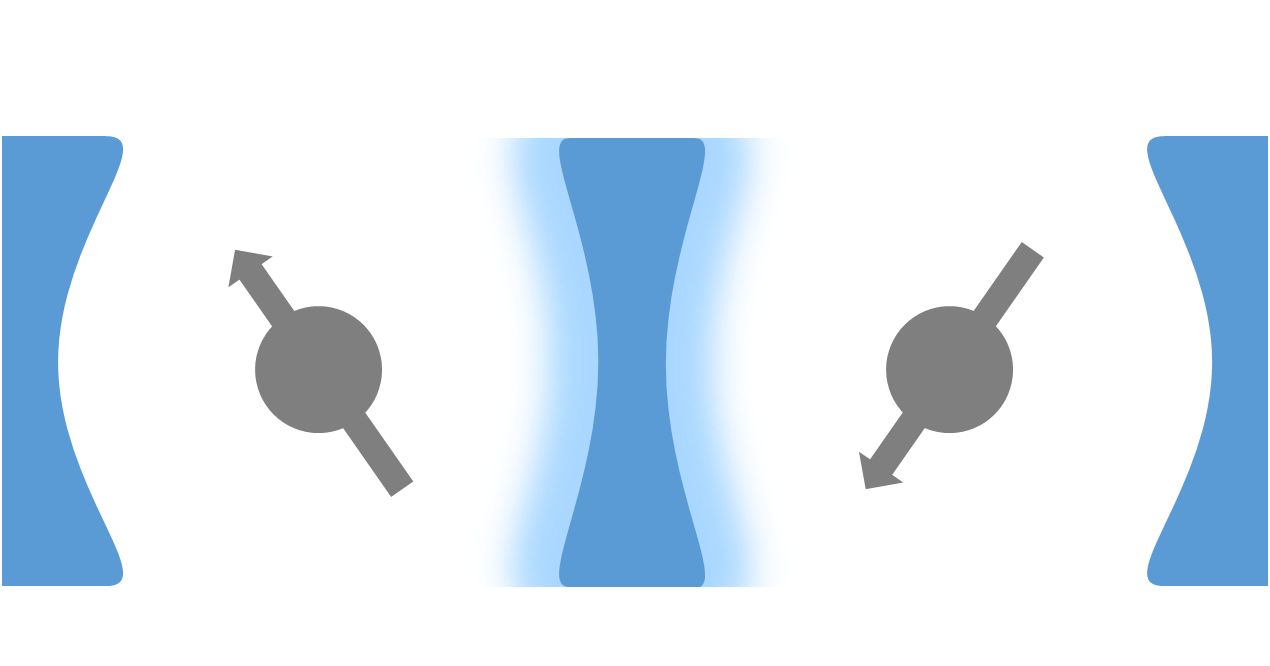}
\caption{Pictorial illustration of the system from a quantum-optical perspective. Two single-mode cavities, each one containing a qubit, are sharing a partially reflecting mirror. When such a mirror moves relativistically, the generation of DCE
photons into the cavities can highly entangle the two subsystems. }
\label{qo_system}
\end{figure}

Let us first describe our model from a quantum-optical perspective, as shown in Fig. \ref{qo_system}, and then in the framework of superconducting circuits. The system comprises two qubit-cavity subsystems interacting via a partially reflecting mirror shared by the cavities \cite{Plenio2006}. A mechanical oscillation of the shared mirror modulates the boundary condition of the electromagnetic field, leading to a production of DCE photons into the
cavities when such motion happens relativistically \cite{Moore1970}. In the framework of superconducting circuits, the cavities can be implemented by coplanar waveguide resonators \cite{Wallraff2008}, which can be coupled to superconducting qubits built from Josephson junctions (JJs) \cite{Clarke2008}. The electrical device equivalent to the shared mirror \cite{Wendin2006} can be implemented through a superconducting quantum interference device (SQUID) \cite{Delin1991}, a superconducting loop interrupted by two JJs and threaded by an external magnetic flux. The latter can be used to fast modulate the interaction between the resonators and their boundary conditions, as a moving mirror \cite{Simulation2014}, provided the modulation frequency is smaller than the SQUID plasma frequency~\cite{CMWilson2011}. Moreover, in order to remain with well-defined cavity modes, the amplitude of deviation in the effective cavity length, induced by the modulation of the boundary conditions, must be smaller than the effective length itself \cite{Felicetti2014}.

The derivation of the Hamiltonian describing the system of Fig.~\ref{qo_system} in the realm of superconducting circuits is shown in Ref.~\cite{Felicetti2014}. The unitary dynamics is given by ($\hbar =1$)~\cite{Fel2014Sup} 
\begin{align}
\mathcal{H}=& \sum_{\mu =1}^{2}H_{R}^{\mu }+\sum_{\mu =1}^{2}\alpha _{\mu}\cos (\omega _{d}t) (a_{\mu }^{\dag }+a_{\mu })^{2}\notag \\
& +\alpha _{0}\cos (\omega _{d}t) (a_{1}^{\dag}+a_{1}) (a_{2}^{\dag }+a_{2}) ,  \label{H_total}
\end{align}
with the quantum Rabi Hamiltonian,
\begin{equation}
H_{R}^{\mu }=\omega _{\mu }a_{\mu }^{\dag }a_{\mu }+\frac{\omega _{q}^{\mu }}{2}\sigma _{z}^{\mu }+g_{\mu }\sigma _{x}^{\mu }(a_{\mu }^{\dag}+a_{\mu }),
\label{H_Rabi}
\end{equation}
describing the interaction between the qubit $\mu $ and its corresponding resonator. Here, $a_{\mu }^{\dag }$ $\left(a_{\mu }\right) $ is the creation (annihilation) operator of the mode of the resonator field with frequency $\omega _{\mu }$, $\sigma _{z,x}^{\mu }$ are the Pauli operators of the qubit with transition frequency $\omega_{q}^{\mu }$, and the qubit-resonator coupling strength is denoted by $ g_{\mu}$. The SQUID, which is modulated by an external flux at frequency $\omega _{d}$, induces two time-dependent processes in the dynamics: a production of unwanted photons in each resonator due to an individual parametric driving field with strength $\alpha _{\mu }$ (single-mode squeezing), and a coupling between the field quadratures of the resonators with strength $\alpha _{0}=2\sqrt{\alpha _{1}\alpha _{2}}$.

\section{Generation of Entanglement}

\subsection{Strong coupling regime}

When each qubit-resonator subsystem satisfies the condition $g_{\mu }\ll \omega_{\mu }\sim \omega _{q}^{\mu }$, we can make use of the rotating-wave approximation (RWA) to neglect the counter-rotating terms in the Rabi Hamiltonian, since they oscillate very rapidly. Therefore, Eq.~(\ref{H_Rabi}) can be\ replaced by the Jaynes-Cummings (JC) Hamiltonian,
\begin{equation}
H_{\text{JC}}^{\mu }=\omega _{\mu }a_{\mu }^{\dag }a_{\mu }+\frac{\omega
_{q}^{\mu }}{2}\sigma _{z}^{\mu }+g_{\mu }(\sigma _{-}^{\mu }a_{\mu
}^{\dag }+\sigma _{+}^{\mu }a_{\mu }) , 
\label{H_JC}
\end{equation}
where $\sigma _{\pm }^{\mu }$ are the raising and lowering operators of the qubit $\mu $. Equation~(\ref{H_JC}) is typically associated with the SC regime, where the coupling strength is larger than all decoherence rates. When this condition is not met, we are dealing with the weak coupling regime, a limit that is not part of this present study.

The control over the tunable driving frequency $\omega _{d}$ allows us to select the desired interaction terms in the system dynamics. For instance, by considering nondegenerate resonators and setting $\omega _{d}=\omega _{1}+\omega _{2}$, the terms induced by the SQUID in Eq. (\ref{H_total}), which are time-independent in the interaction picture, are $\alpha_{0}(a_{1}^{\dag}a_{2}^{\dag} + a_{1}a_{2})/2$. Then, as long as $\alpha _{\mu },\alpha _{0}\ll \omega _{\mu},\vert \omega _{1}-\omega _{2}\vert$, the remaining time-dependent terms are fast oscillating and can be neglected by means of RWA. In such circumstances, the dynamics is given by
\begin{equation}
\mathcal{H}_{\text{SC}}=\sum_{\mu =1}^{2}H_{\text{JC}}^{\mu }+\frac{\alpha
_{0}}{2} (a_{1}^{\dag }a_{2}^{\dag }e^{-i\omega
_{d}t}+a_{1}a_{2}e^{i\omega _{d}t}).  \label{H_SC}
\end{equation}

\begin{figure}[b]
\includegraphics[width=0.45\textwidth]{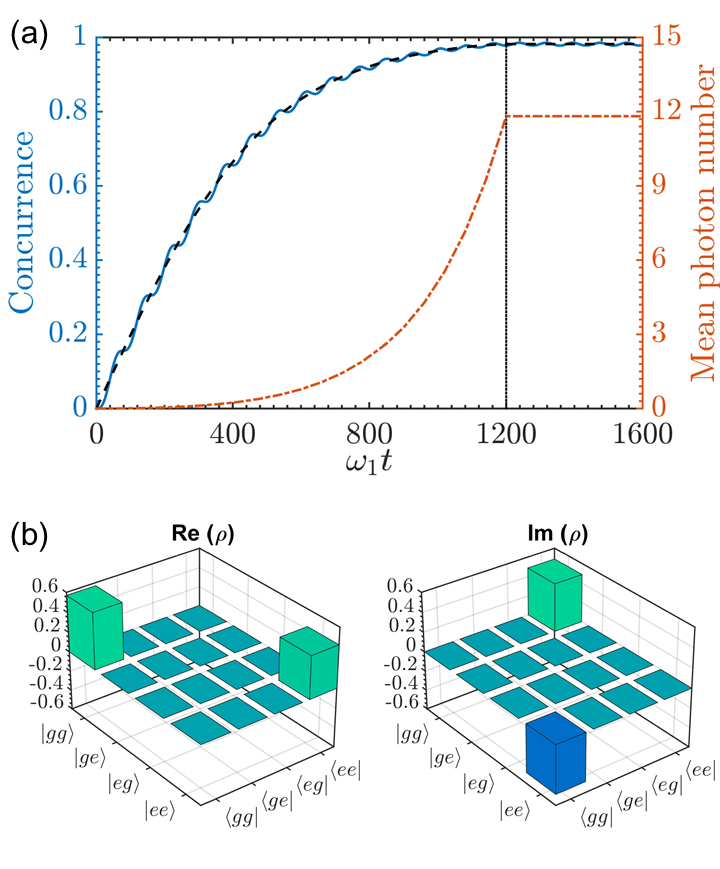}
\caption{(a) Time evolution of concurrence between the two qubits (solid line) and mean photon number of the first resonator (dashed-dotted line). Each qubit is resonantly coupled to its corresponding resonator with the same interaction strength $g_{0}=g_{1}=g_{2}=0.04\protect \omega _{1}$, with $\protect\omega _{2}=1.25\protect\omega _{1}$ and $ \protect\alpha _{0}=0.1g_{0}$. The vertical dotted line represents the time, $\omega_{1}t_{
\text{SO}}=1200$, when the SQUID external flux is switched off. The dashed line is the concurrence computed using Eq. (\protect\ref{H_SC3}). (b)~Real and imaginary parts of the reduced density matrix of the two-qubit system at $t_{\text{SO}}$. The semi-analytical model developed in this work agrees with our numerical studies, some of which were previously presented in Ref.~\cite{Felicetti2014}.}
\label{conc_sc_ideal}
\end{figure}

In this Hamiltonian, the two-mode squeezing term proportional to $\alpha _{0}$ generates pairs of entangled photons shared by the resonators, where such entanglement may be transferred to the qubits via the
JC interaction. Indeed, as shown in Ref. \cite{Felicetti2014}, it is possible to achieve\ maximal entanglement (Bell state) between the qubits. In Fig. \ref{conc_sc_ideal}, we show the results of the
protocol consisting of cooling down the system to its ground state, turning the external flux on and switching it off at time $t_{\text{SO}}$ when the maximal entangled state is reached~\cite{Felicetti2014}. In Fig. \ref{conc_sc_ideal}(a), we see from the time evolution of the concurrence \cite{Wootters1998} that the qubits reach an almost maximally entangled state ($\mathcal{C}_{\text{SC}}\sim 0.98$), which is very close to the Bell state $\vert \psi \rangle = (\vert \text{ee}\rangle +i\vert \text{gg}\rangle) /\sqrt{2}$ [Fig. \ref{conc_sc_ideal}(b)]. If we consider $\omega _{1}/2\pi =5$ GHz, we have $t_{\text{SO}}\sim 40$ ns for the parameters used in Fig. \ref{conc_sc_ideal}. However, $t_{\text{SO}} \approx 10$$-$$500$ ns for a wide range of realistic system parameters \cite{Felicetti2014, Fel2014Sup}. It is worth to stress that $\mathcal{C}_{\text{SC}}$ converges to unity if we consider a larger $t_{\text{SO}}$, i.e., the qubits asymptotically reach a maximally entangled state. Consequently, they are not entangled with the resonator modes, even if they strongly interact during the whole time evolution. However, aside from the entanglement between the qubits in the SC regime, there is polaritonic entanglement, because the two-resonator subsystems are also entangled asymptotically. In the following, we provide an intuitive explanation for this non-trivial dynamics.

In order to examine the dynamics in more detail, let us rewrite Eq. (\ref{H_SC}) in the eigenbasis of JC Hamiltonian. As we assume that the whole system is initially prepared in the ground state, $
\vert \phi _{0}\rangle \equiv \vert \text{g},0\rangle\otimes \vert \text{g},0\rangle $, its dynamics is restricted to the Hilbert subspace spanned by $\{\vert \phi _{0}\rangle,\vert \phi _{n}^{\pm }\rangle \equiv \vert \pm,n\rangle \otimes \vert \pm ,n\rangle ,\vert \Psi
_{n}\rangle \equiv ( \vert \text{g},n\rangle \otimes\vert \text{g},n\rangle -\vert \text{e},n-1\rangle\otimes \vert \text{e},n-1\rangle) /\sqrt{2}\}$, with $\vert \pm ,n\rangle=(\vert \text{g},n\rangle\pm \vert \text{e},n-1\rangle)/\sqrt{2}$ (for $n=1,2,3$ $...$). This results in a gain in the required computational effort, since the dimension of this subspace grows linearly with $n_{\text{max}}$ instead of quadratically, where $n_{\text{max} }$ is the maximum occupation number considered in the Fock basis of the cavity modes. Here, setting $\omega _{\mu }=\omega _{q}^{\mu }$ and $g_{0}=g_{1}=g_{2}$, we have for both subsystems $H_{\text{JC}}^{\mu }\vert \pm ,n\rangle _{\mu}=\mathcal{E}_{\pm ,n}^{\mu }\vert \pm ,n\rangle _{\mu }$ with $\mathcal{E}_{\pm ,n}^{\mu }=(n-1/2) \omega _{\mu }\pm g_{0}\sqrt{n}$, whereas $\mathcal{E}_{0}^{\mu }=-\omega _{\mu }/2$ for the ground state $\vert \text{g},0\rangle _{\mu }$. Thus, defining $\varpi =\omega _{1}+\omega _{2}$, Eq. (\ref{H_SC}) can be exactly rewritten as
\begin{align}
\mathcal{H}_{\text{SC}}=& -\frac{\varpi }{2}\vert \phi_{0}\rangle \langle \phi _{0}\vert  \notag \\
& +\sum_{n=1}^{\infty }\left[ \left(n-\frac{1}{2}\right) \varpi \right]\vert \Psi _{n}\rangle \langle \Psi _{n}\vert  \notag \\
& +\sum_{n=1}^{\infty }\left[ \left( n-\frac{1}{2}\right) \varpi +2g_{0} \sqrt{n}\right] \vert \phi _{n}^{+}\rangle \langle \phi_{n}^{+}\vert  \notag \\
& +\sum_{n=1}^{\infty }\left[ \left( n-\frac{1}{2}\right) \varpi -2g_{0}\sqrt{n}\right] \vert \phi _{n}^{-}\rangle \langle \phi_{n}^{-}\vert  \notag \\
& +\frac{\alpha _{0}}{4}(\mathcal{O}_{0}e^{i\varpi t}+\mathcal{O}_{0}^{\dag }e^{-i\varpi t})  \notag \\
& +\frac{\alpha _{0}}{8}\sum_{n=1}^{\infty }(\mathcal{O}_{n}e^{i\varpi t}+\mathcal{O}_{n}^{\dag }e^{-i\varpi t}) ,
\label{H_SC2}
\end{align}
with
\begin{equation}
\mathcal{O}_{0}=\vert \phi _{0}\rangle (\langle \phi_{1}^{+}\vert +\langle \phi _{1}^{-}\vert) +\sqrt{2}\vert \phi _{0}\rangle \langle \Psi _{1}\vert ,
\label{O0}
\end{equation}
and
\begin{eqnarray}
\mathcal{O}_{n}&=&(\sqrt{n+1}+\sqrt{n})^{2}(\vert\phi _{n}^{+}\rangle \langle \phi _{n+1}^{+}\vert+\vert \phi _{n}^{-}\rangle \langle \phi_{n+1}^{-}\vert)  \notag \\
&&+(\sqrt{n+1}-\sqrt{n})^{2}(\vert \phi_{n}^{+}\rangle \langle \phi _{n+1}^{-}\vert +\vert\phi _{n}^{-}\rangle \langle \phi _{n+1}^{+}\vert) \notag \\
&&+\sqrt{2}(\vert \phi _{n}^{+}\rangle +\vert \phi_{n}^{-}\rangle) \langle \Psi _{n+1}\vert  \notag \\
&&+\sqrt{2}\vert \Psi _{n}\rangle(\langle \phi_{n+1}^{+}\vert +\langle \phi _{n+1}^{-}\vert)\notag \\
&&+2( 2n+1) \vert \Psi _{n}\rangle \langle \Psi_{n+1}\vert .  \label{On}
\end{eqnarray}

\begin{figure}[t]
\includegraphics[width=0.45\textwidth]{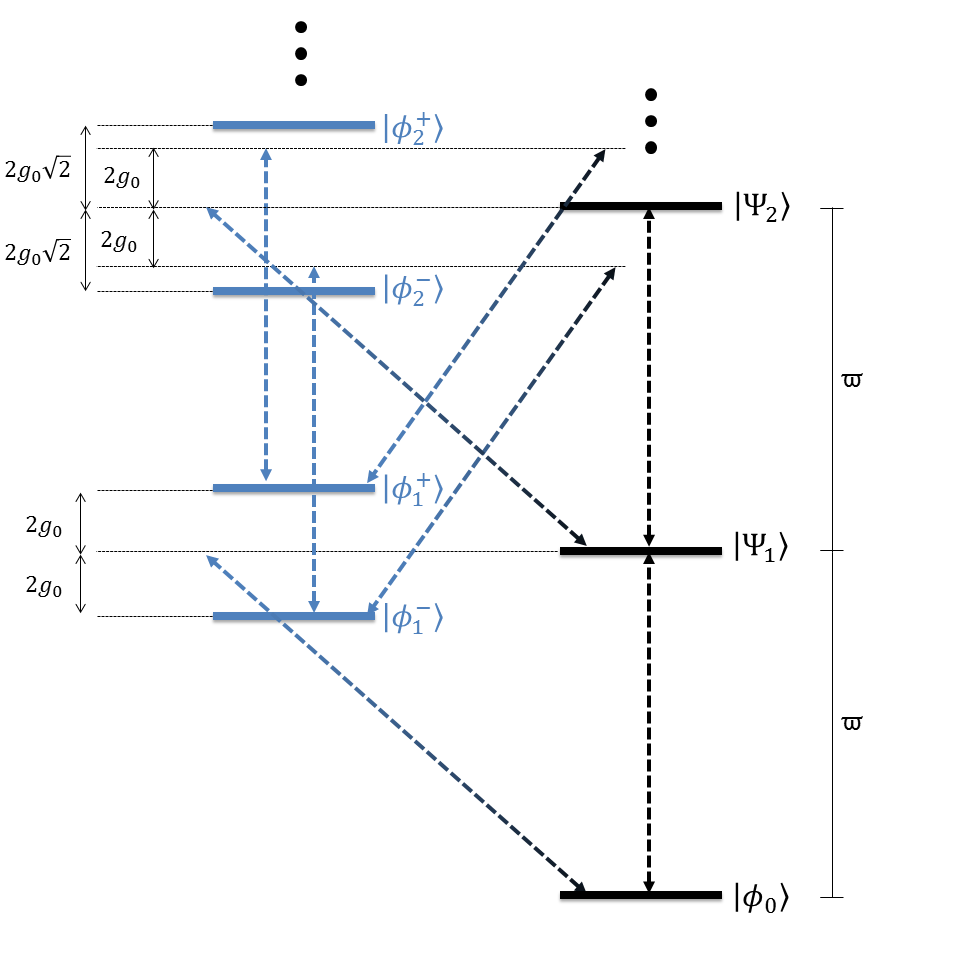}
\caption{Energy-level diagram of the Hamiltonian given by Eq. (\protect\ref{H_SC2}), which consists in a harmonic ladder and an anhamonic part. The dashed arrows stand for the transitions induced by the two-mode squeezing term, where only the transitions between adjacent states of the harmonic ladder are induced resonantly.}
\label{ladder}
\end{figure}

It is straightforward to observe from Eq. (\ref{H_SC2}) that the system energy spectrum consists in a harmonic ladder with fundamental frequency $\varpi $ given by the subspace $\{\vert \phi _{0}\rangle,\vert \Psi _{n}\rangle\} $, and an anharmonic part due to the doublets $\{\vert \phi _{n}^{+}\rangle ,\vert \phi_{n}^{-}\rangle\} $. Furthermore, as depicted in Fig. \ref{ladder}, the two-mode squeezing term induces resonant transitions only between adjacent states of the harmonic part, since we are considering $\omega_{d}=\varpi $. Hence, as long as $\alpha _{0}\ll g_{0}$, the fast-oscillating terms in Eq. (\ref{H_SC2}) coming from the non-resonantly induced transitions can be neglected due to RWA. Then, in the interaction picture with respect to $H_{0}=H_{\text{JC}}^{1}+H_{\text{JC}}^{2}$, the system dynamics is effectively described by 
\begin{eqnarray}
\mathcal{H}_{\text{SC}}^{I\text{(eff)}} &\approx &\frac{\alpha _{0}}{2}\left[\sum_{n=1}^{\infty }\frac{(2n+1) }{2}(\vert \Psi_{n}\rangle \langle \Psi _{n+1}\vert +\vert \Psi_{n+1}\rangle \langle \Psi _{n}\vert) \right.  \notag \\
&&+\left. \frac{1}{\sqrt{2}}(\vert \phi _{0}\rangle\langle \Psi _{1}\vert +\vert \Psi _{1}\rangle\langle \phi _{0}\vert) \right] . \label{H_SC3}
\end{eqnarray}

The dashed line in Fig.~\ref{conc_sc_ideal}(a) represents the concurrence computed via Eq. (\ref{H_SC3}), which is in good agreement with the results obtained using the Hamiltonian without approximations (solid line). The small oscillations in its evolution are remnants of the nonresonant terms discarded in Eq. (\ref{H_SC3}), that is, there is a deviation from the harmonic dynamics due to the adiabatic elimination \cite{Sanz2015}, but the latter still prevails, mainly for $t\gtrsim 1/\alpha _{0}$. The effective Hamiltonian in Eq. (\ref{H_SC3}) provides an intuitive explanation for the entanglement generation depicted in Fig.~\ref{conc_sc_ideal}. The set of coupled differential equations provided by Eq. (\ref{H_SC3}) has a uniform three-term recurrence relation, which can be semi-analytically solved using Fourier transform with techniques based on continued fractions \cite{Gautschi1981,Gonoskov2014}. For the sake of brevity, we are not going to discuss this solution here, since its mathematical expression does not give more insights on the system dynamics.

Dissipative processes usually destroy quantum coherence such as entanglement. For the protocol discussed above, spontaneous emission of the qubits leaks their shared information gradually into the environment, while a dephasing process destroys the coherence of quantum superpositions, decreasing the entanglement with time. Likewise, the leaking of photons out of the resonators leads to the same effect, since the qubits share information through the DCE-generated pairs of correlated photons. From another perspective, the system dynamics is no longer kept in the harmonic subspace, since such dissipative effects, together with the incoherently induced transitions, drive the qubits to a mixed state, which decays to the ground state after switching off the external flux. 

\begin{figure}[b]
\includegraphics[width=0.45\textwidth]{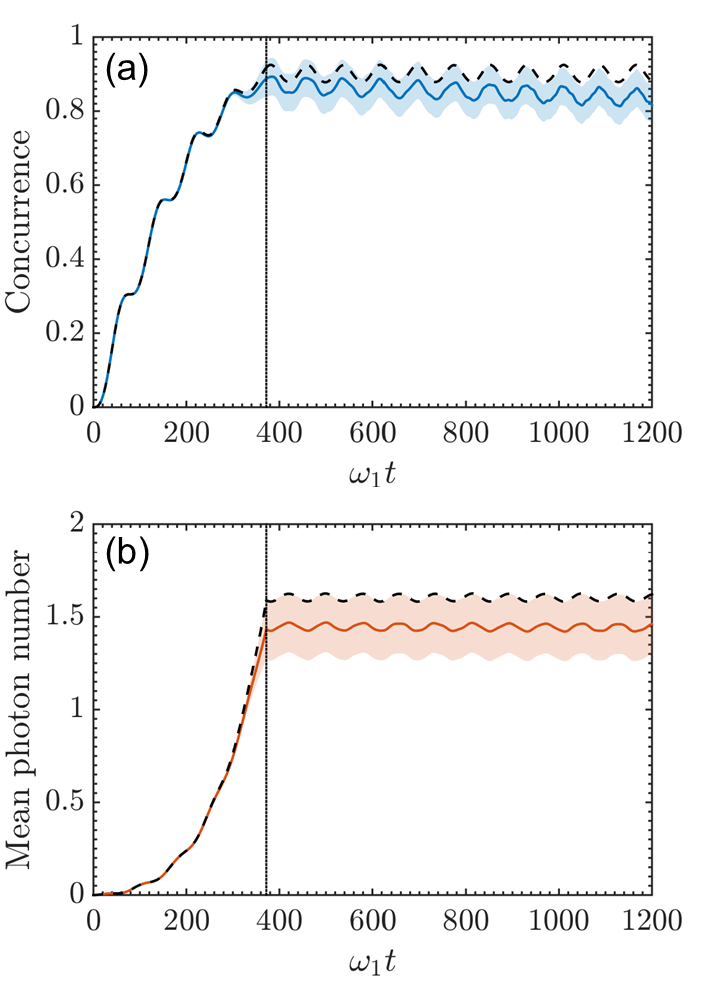}
\caption{Time evolution of (a) the concurrence and (b)~the mean photon number when there is dissipation (solid lines) in comparison to the case when there is not (dashed lines). Here, we consider $\protect\kappa _{\protect\mu }=\protect\gamma _{\protect\mu }=\protect\gamma _{\protect\mu }^{\text{ph}}=5\times 10^{-6}\protect\omega _{1}$ and the same parameters of Fig. \protect\ref{conc_sc_ideal}, except $t_{\text{SO}}$ and $\protect\alpha _{0}=0.2g_{0}$. The shaded area stands for the numerical errors because of the truncation of the Fock basis describing the resonators.}
\label{conc_sc_diss}
\end{figure}

To test the sensitivity of the generated entanglement with respect to such dissipative effects, the non-unitary dynamics of the system can be described by the standard quantum optics master equation~\cite{Blais2011,Breuer}, provided that each qubit-resonator subsystem is described by the JC model,
\begin{eqnarray}
\dot{\rho} &=&-i[\mathcal{H}_{\text{SC}},\rho] +\sum_{\mu=1}^{2}\kappa _{\mu }\mathcal{D}[a_{\mu }] \rho  \notag \\
&&+\sum_{\mu =1}^{2}( \gamma _{\mu }\mathcal{D}[ \sigma _{-}^{\mu}] \rho +\gamma _{\mu }^{\text{ph}}\mathcal{D}[ \sigma _{z}^{\mu }] \rho) , 
\label{ME_SC}
\end{eqnarray}
where $\kappa _{\mu }$ is the decay rate of the mode of the resonator $\mu $, $\gamma _{\mu }$ and $\gamma _{\mu }^{\text{ph}}$ are the relaxation and the pure dephasing rates of the qubit $\mu $, respectively, and $\mathcal{D} [\mathcal{O}] \rho =(2\mathcal{O}\rho \mathcal{O}^{\dag }-
\mathcal{O}^{\dag }\mathcal{O}\rho -\rho \mathcal{O}^{\dag }\mathcal{O})/2$. In Fig. \ref{conc_sc_diss}, we show similar results to those of Fig. \ref{conc_sc_ideal}(a), considering identical and realistic dissipation rates $\kappa _{\mu }=\gamma _{\mu }=\gamma _{\mu }^{\text{ph}}=5\times 10^{-6}\omega _{1}$ \cite{Wallraff2008,Chang2013}, with $\alpha_{0}=0.2g_{0}$, but also assuming a shorter $t_{\text{SO}}$ ($\sim$$12$ ns when $\omega _{1}/2\pi =5$ GHz). The reason is that the numerical resolution of Eq. (\ref{ME_SC}) is a hard computational task, since the photon mean number grows exponentially at short times until its saturation. At the same time, the Liouvillian space grows with $n_{\max}^{4}$ instead of the $n_{\max }^{2}$ dependence of the full Hilbert space. However, even in this adverse technical situation, we note that it is still possible to achieve a reasonable value of entanglement in the presence of dissipation ($\mathcal{C}_{\text{SC}}\sim 0.89$) in comparison with the ideal case for the same $t_{\text{SO}}$ ($\mathcal{C}_{\text{SC}}\sim 0.92$). Therefore, we also theoretically demonstrate the robustness of the entanglement generation against the dissipative processes with realistic decay rates. The shaded area stands for the numerical errors, because of the truncation of the Fock basis describing the resonators with insufficient numbers of photons.

\begin{figure}[t]
\includegraphics[width=0.45\textwidth]{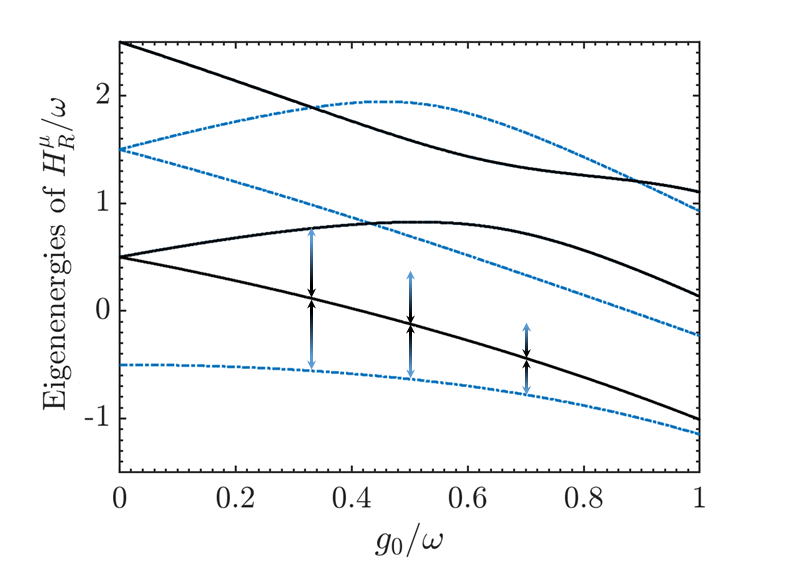}
\caption{Energy spectrum of the quantum Rabi model with respect to the ratio $g_{0}/\protect\omega $, where $\protect\omega $ is the frequency of both resonator mode and qubit, and $g_{0}$ is the coupling strength of the qubit-resonator interaction. The dash-dotted and solid lines stand for the eigenstates with positive and negative parity, respectively. The arrows indicate the energy gap between the ground and first excited states for $g_{0}/\protect\omega=(0.25,0.5,0.75)$. If we resonantly induce a transition from the ground to the first excited state by a single photon process, the probability of inducing a transition from the first excited state to the next allowed one (next dash-dotted line) by the same process is very small due to nonresonance. Then, the strong anharmonicity of the energy spectrum in the USC regime ($0.1\lesssim g_{0}/\protect\omega <1$) allows us to work in a low-energy subspace, that is, the ground and the first excited state of a quantum Rabi system in the USC regime can be used as a qubit~\protect\cite{Nataf2011}.}
\label{spec_rabi}
\end{figure}

\subsection{Ultrastrong coupling}

For the case in which both qubit-resonator subsystems are in the USC regime, $g_{\mu }\lesssim \max ( \omega _{\mu },\omega _{q}^{\mu }) $, the RWA is no longer justified, i.e., the quantum Rabi model has to be used for describing the dynamics of each subsystem. Moreover, neither the degrees of freedom of the superconducting qubit, nor the ones of its resonator, can be individually addressed in the USC regime~\cite{Ridolfo2012} (as well as in the DSC regime), since now they are part of a merged dressed qubit-cavity system. In this context, the motion of the shared mirror no longer changes the boundary conditions of the electromagnetic field of the cavities, but now it changes the boundary conditions of the normal eigenmodes of each hybrid system. Therefore, in the case of SC, the DCE generates pairs of entangled photons into the cavities which can be absorbed by the superconducting qubits due to the qubit-cavity coupling. While in the case of USC (and DSC), the whole light-matter system is directly affected by the oscillations of the mirror. Thus, because of the insolubility of the hybrid systems, we focus on the entanglement between polaritons (qubit-resonator collective excitations) instead of the entanglement between the original superconducting qubits.

\begin{figure}[b]
\includegraphics[width=0.45\textwidth]{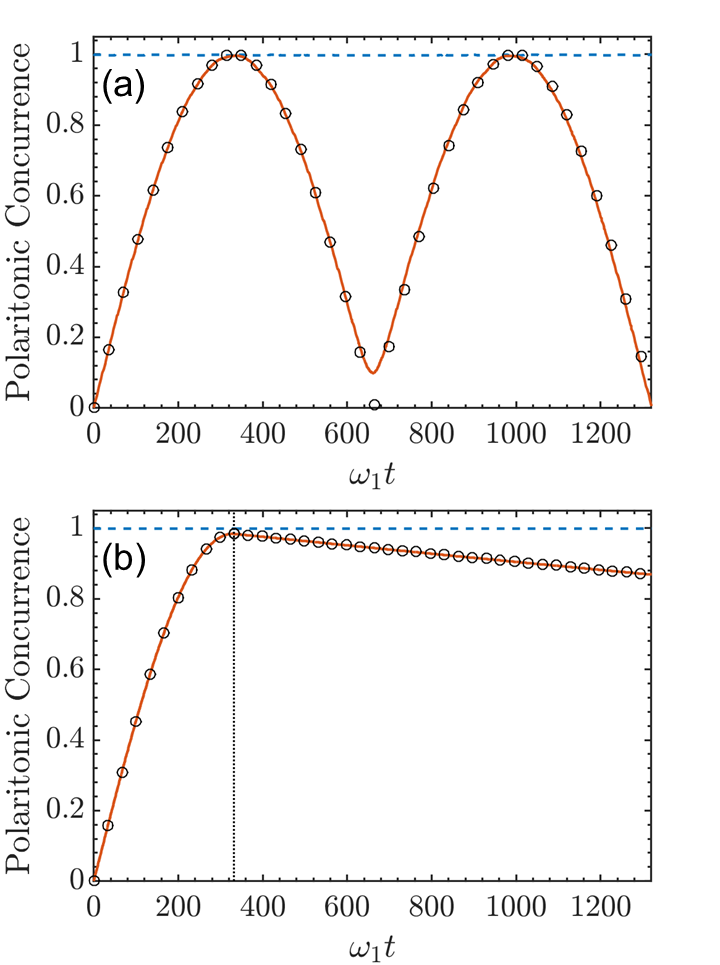}
\caption{(a) Time evolution of the concurrence (solid line) between two identical quantum Rabi systems computed using the Hamiltonian without approximations, Eq. (\protect\ref{H_USC}), considering $g_{0}=0.15\protect\omega $ and $\protect\alpha_{0}=0.05g_{0}$. The circles represent the concurrence computed via Eq. (\protect\ref{H_USC2}), $\mathcal{C}=\vert \sin ( \protect\lambda t) \vert $, which is in good agreement with the exact numerical solution. Let $P( \vert \protect\varphi _{k}^{\protect\mu}\rangle)$ the probability of the subsystem $\protect\mu $ being in the eigenstate $\vert \protect\varphi _{k}^{\protect\mu}\rangle $,\ the dashed line stands for $P(\vert \protect\varphi _{0}^{1}\rangle) +P(\vert \protect\varphi_{1}^{1}\rangle) $, which shows us the validity of the assumption that each quantum Rabi subsystem in the USC regime works as a qubit since we have $P(\vert \protect\varphi _{0}^{1}\rangle)+P(\vert \protect\varphi _{1}^{1}\rangle) \sim 1$ during the evolution. (b) Similar to (a) in the case when dissipative processes take place. The solid line and the circles refer to the solution of Eq. (\protect\ref{ME_USC}) and Eqs. (\protect\ref{Conc_diss})$-$(\protect\ref{C_USC_tsoplus}), respectively, considering $\protect\kappa =\protect\gamma =0.5\times 10^{-4}\protect\omega $ and $t_{\text{SO}}=\protect\pi /2\protect\lambda $ (dotted line).}
\label{concUSC}
\end{figure}

As in the previous subsection, we rewrite Eq. (\ref{H_total}) in the eigenbasis of the quantum Rabi Hamiltonian of Eq.~(\ref{H_Rabi}), and then select the desired interaction terms in the system dynamics by adjusting the driving frequency $\omega _{d}$ of the SQUID external flux. Consider $\vert \varphi _{k}^{\mu }\rangle $ an eigenstate of $H_{R}^{\mu }$, so that $H_{R}^{\mu }\vert \varphi _{k}^{\mu}\rangle =\mathcal{\epsilon }_{k}^{\mu }\vert \varphi _{k}^{\mu}\rangle $. A quantum Rabi system in the USC regime\ has a strong anharmonicity in its energy spectrum which allows us to work in a low-energy subspace (see Fig. \ref{spec_rabi}). This permits to focus only on the transition between the ground and the first excited polaritonic states of each quantum Rabi system, which always have opposite parity $\Pi _{\mu }=-\sigma _{z}^{\mu }e^{i\pi a_{\mu }^{\dag }a_{\mu }}$ due to the discrete $\mathbb{Z}_{2}$-symmetry \cite{Braak2011}. Given that $a_{\mu }^{2}$ and $a_{\mu}^{\dag 2}$ leave the parity $\Pi _{\mu }$ of a given state untouched \cite{Felicetti2015}, we have $\langle \varphi _{k^{\prime }}^{\mu}\vert a_{\mu }^{2}\vert \varphi _{k}^{\mu }\rangle =0$ and $\langle \varphi _{k^{\prime }}^{\mu }\vert a_{\mu }^{\dag2}\vert \varphi _{k}^{\mu }\rangle =0$ for any pair of states with different parity. This means that, in opposition to the case discussed in the preceding subsection and in Ref.~\cite{Felicetti2014}, we do not have to consider nondegenerate resonators in order to get rid of single-mode squeezing terms. Thus, we can write Eq. (\ref{H_total}) in the interaction picture with respect to $H_{0}=H_{R}^{1}+H_{R}^{2}$ as
\begin{eqnarray}
\mathcal{H}_{\text{USC}}^{I} &=&\alpha _{0}\cos ( \omega _{d}t)\sum_{\substack{ k,l  \\ k\neq l}}\sum_{\substack{ \eta ,\nu  \\ \eta \neq \nu }}X_{1}^{kl}X_{2}^{\eta \nu }\vert \varphi _{k}^{1}\rangle
\langle \varphi _{l}^{1}\vert  \notag \\
&&\otimes \vert \varphi _{\eta }^{2}\rangle \langle \varphi_{\nu }^{2}\vert e^{i\Delta _{kl}^{1}}e^{i\Delta _{\eta \nu }^{2}},
\label{H_USC}
\end{eqnarray}
where $X_{\mu }^{kl}=\langle \varphi _{k}^{\mu }\vert (a_{\mu }+a_{\mu }^{\dag }) \vert \varphi _{l}^{\mu }\rangle $ and $\Delta _{kl}^{\mu }=\mathcal{\epsilon }_{k}^{\mu }-\mathcal{\epsilon }_{l}^{\mu }$.

If we set $\omega _{d}=\Delta _{01}^{1}+\Delta _{01}^{2}$, we can neglect fast-oscillating terms in Eq. (\ref{H_USC}) by means of RWA as long as $\alpha _{0}\vert X_{1}^{kl}X_{2}^{\eta \nu }\vert \ll \vert\omega _{d}\pm ( \Delta _{kl}^{1}+\Delta _{\eta \nu }^{2})\vert $, so that
\begin{equation}
\mathcal{H}_{\text{USC}}^{I\text{(eff)}}\approx \frac{\lambda }{2}(\mathcal{S}_{-}^{1}\mathcal{S}_{-}^{2}+\mathcal{S}_{+}^{1}\mathcal{S}_{+}^{2}) ,  \label{H_USC2}
\end{equation}
with $\lambda =\alpha _{0}X_{1}^{01}X_{2}^{01}$ and $\mathcal{S}_{-}^{\mu }=\mathcal{S}_{+}^{\mu \dag }=\vert \varphi _{0}^{\mu }\rangle\langle \varphi _{1}^{\mu }\vert $.

It has been shown in Ref.~\cite{Nataf2011} that the ground and first excited levels of a quantum Rabi system in USC form a protected qubit robust against a general class of anisotropic environments, i.e., when they are coupled to the original superconducting qubit in a different direction to the qubit-resonator coupling direction. Here, we observe from Eq. (\ref{H_USC2}) that the DCE induces an interaction between such protected qubits (quantum XY spin interaction) which leads to a polaritonic concurrence 
\begin{equation}
\mathcal{C}_{\text{USC}}(t) =\vert \sin ( \lambda t)\vert ,
\label{Conc_ideal}
\end{equation}
if the whole system is initially in the ground state $\vert \varphi_{0}^{1}\rangle \vert \varphi _{0}^{2}\rangle $, as displayed in Fig. \ref{concUSC}(a). Hence, a maximally entangled polaritonic
state between hybrid systems, $(\vert \varphi_{0}^{1}\rangle \vert \varphi _{0}^{2}\rangle -i\vert\varphi _{1}^{1}\rangle \vert \varphi _{1}^{2}\rangle) /\sqrt{2}$, can be reached switching the external flux off at $\lambda t_{\text{SO}}\sim \pi /2$, which is $t_{\text{SO}}\sim 10$ ns considering $\omega _{1}=5$ GHz.

For systems described by the quantum Rabi model in the USC regime, wherein the RWA breaks down, the standard quantum optics master equation no longer describes properly the system dynamics, so that a master equation considering the qubit-resonator coupling is required \cite{Blais2011}. Although the qubit built from the two lowest eigenstates of a Rabi system in USC is robust against anisotropic noise sources, it is not against either the qubit noise in the direction of the coupling between the resonator and the original superconducting qubit, or the noise associated to the resonator~\cite{Nataf2011}. By following the approach of Ref. \cite{Blais2011} and
considering only noises which such new qubits are not protected, the dynamics of the system is governed by the master equation 
\begin{align}
\dot{\rho}=& -i[ \mathcal{H}_{\text{USC}}^{I},\rho]  \notag \\
& +\sum_{\mu =1}^{2}\sum_{k,l>k}(\Gamma _{\kappa _{\mu}}^{kl}+\Gamma _{\gamma _{\mu }}^{kl}) \mathcal{D}[ \vert\varphi _{k}^{\mu }\rangle \langle \varphi _{l}^{\mu }\vert] , 
\label{ME_USC}
\end{align}
where the resonator and the qubit baths induce transitions from $ \vert \varphi _{l}^{\mu } \rangle \ $to $ \vert \varphi_{k}^{\mu } \rangle$ at rates $\Gamma _{\kappa _{\mu }}^{kl}=\kappa _{\mu } ( \Delta _{kl}^{\mu} )  \vert X_{\mu }^{kl} \vert ^{2}$ and $\Gamma_{\gamma_{\mu }}^{kl}=\gamma _{\mu } ( \Delta _{kl}^{\mu } )  \vert \langle \varphi _{k}^{\mu } \vert \sigma _{x}^{\mu } \vert\varphi _{l}^{\mu } \rangle  \vert ^{2}$, respectively. Here, $\kappa _{\mu } ( \Delta _{kl}^{\mu } ) $ and $\gamma _{\mu } (\Delta _{kl}^{\mu } ) $ are rates which are proportional to noise spectra, for resonator and original qubit environments of subsystem $\mu $, respectively.

When each hybrid system can be reduced to a two-level system, the effective master equation is given by 
\begin{equation}
\dot{\rho}=-i [ \mathcal{H}_{\text{USC}}^{I\text{(eff)}},\rho  ] +\sum_{\mu =1}^{2}\Gamma _{\mu }\mathcal{D} [ \mathcal{S}_{-}^{\mu }  ] ,
\label{ME_USC2}
\end{equation}
with $\Gamma _{\mu }= ( \Gamma _{\kappa _{\mu }}^{01}+\Gamma _{\gamma_{\mu }}^{01} ) $. Assuming two identical hybrid systems, Eq. (\ref{ME_USC2}) predicts the following concurrence between them,
\begin{equation}
\mathcal{C}_{\text{USC}} ( t ) =\frac{\lambda }{ ( \Gamma^{2}+\lambda ^{2} ) }\max  [ 0, \vert u ( t ) \vert -p ( t )  ] ,
\label{Conc_diss}
\end{equation}
with
\begin{eqnarray}
u ( t ) &=&\Gamma +e^{-\Gamma t} [ \lambda \sin  ( \lambda t ) -\Gamma \cos  ( \lambda t )  ] ,  \label{u} \\
p ( t ) &=&e^{-\Gamma t} [ \lambda \sinh  ( \Gamma t ) -\Gamma \sin  ( \lambda t )  ] .  \label{p}
\end{eqnarray}
If $\Gamma \ll \lambda $, we have, for $t_{\text{SO}}\sim \pi /2\lambda$,
\begin{equation}
\mathcal{C}_{\text{USC}} ( t_{\text{SO}} ) \sim 1- ( \pi -2 ) \frac{\Gamma }{\lambda }.  \label{C_USC_tso}
\end{equation}
And for $t>t_{\text{SO}}$ ($\alpha _{0}=0 \rightarrow \lambda =0$), we have 
\begin{equation}
\mathcal{C}_{\text{USC}} ( t>t_{\text{SO}} ) =\mathcal{C}_{\text{USC}} ( t_{\text{SO}} ) e^{-2\Gamma t}.  \label{C_USC_tsoplus}
\end{equation}
Thus, the dissipative processes reduce the maximum concurrence of the ideal case in $\delta = ( \pi -2 ) \Gamma /\lambda $ at $t_{\text{SO}}\sim \pi /2\lambda $, and then the entanglement decreases exponentially at a rate of $2\Gamma $, as depicted in Fig. \ref{concUSC}(b).

\subsection{Deep strong coupling}

In the DSC regime, $g_{\mu }>\max  ( \omega _{\mu },\omega _{q}^{\mu} ) $, the lowest eigenstates of Eq. (\ref{H_Rabi}) can be analytically obtained by using an adiabatic approximation
approach in the displaced oscillator basis \cite{Irish2005}. It is worth to stress that the greater $g_{\mu }/\max  ( \omega _{\mu },\omega _{q}^{\mu} )$ and/or $\omega _{\mu }/\omega _{q}^{\mu}$, the better the system is described by such approximation \cite{Irish2005}, such that
\begin{equation}
H_{R}^{\mu }\approx \sum_{\theta =-,+}\sum_{N=0}^{\infty }E_{\theta ,N}^{\mu} \vert \Lambda _{\theta ,N}^{\mu } \rangle  \langle \Lambda_{\theta ,N}^{\mu } \vert , 
\label{H_Rabi_DSC}
\end{equation}
with 
\begin{eqnarray}
 \vert \Lambda _{\pm ,N}^{\mu } \rangle &=&\frac{1}{\sqrt{2}} (  \vert + \rangle _{\mu }\otimes  \vert N_{+} \rangle _{\mu }\pm  \vert - \rangle _{\mu }\otimes  \vert N_{-} \rangle_{\mu } ) ,  \label{eigstatesDSC} \\
E_{\pm ,N}^{\mu } &=&E_{N}^{\mu }\pm \Omega _{N}^{\mu },  \label{eigenergDSC}
\end{eqnarray}
where $ \vert \pm  \rangle _{\mu }= (  \vert \text{g}  \rangle _{\mu }\pm  \vert \text{e} \rangle _{\mu } ) / \sqrt{2}$ and $ \vert N_{\pm } \rangle _{\mu }=D_{\mu } ( \mp \beta _{\mu } )  \vert N \rangle _{\mu }$ (displaced Fock state). Here, $D_{\mu } ( \beta _{\mu } ) =e^{\beta _{\mu } ( a_{\mu }^{\dag }-a_{\mu } ) }$ is the displacement operator over resonator $\mu $ with displacement amplitude $\beta _{\mu }=g_{\mu }/\omega _{\mu }$, and
\begin{eqnarray}
E_{N}^{\mu } &=&\omega _{\mu } [ N-\beta _{\mu }^{2} ] ,  \label{EN}
\\
\Omega _{N}^{\mu } &=&\frac{\omega _{q}^{\mu }}{2}e^{-2\beta _{\mu}^{2}}L_{N} ( 4\beta _{\mu }^{2} ),  \label{OmegaN}
\end{eqnarray}
where $L_{N}$ is the Laguerre polynomial of order $N$.

Following the same procedure adopted in the previous subsections, we rewrite Eq. (\ref{H_total}) in the basis $\{ \vert \Lambda _{\pm ,N}^{\mu }\rangle\} $, and then we tune conveniently the frequency $\omega _{d}$. Assuming nondegenerate resonators in order to get rid of single-mode squeezing terms, as in the case of SC regime, we have 
\begin{eqnarray}
\mathcal{H}_{\text{DSC}} &=&\sum_{\mu =1}^{2}\sum_{N=0}^{\infty }\sum_{\theta =-,+}E_{\theta ,N}^{\mu }\vert \Lambda _{\theta ,N}^{\mu }\rangle \langle \Lambda _{\theta ,N}^{\mu }\vert   \notag \\
&+&\alpha_0\cos(\omega_dt)\sum_{N=0}^{\infty }\sum_{\theta =-,+}\left[ \sqrt{N+1}\vert \Lambda _{\theta ,N+1}^{1}\rangle \langle \Lambda _{\theta ,N}^{1}\vert \right.   \notag \\
&&\left. -\beta _{1}\vert \Lambda _{\theta ,N}^{1}\rangle \langle \Lambda _{-\theta ,N}^{1}\vert +\text{H.c.}\right]  \notag \\
&\otimes &\sum_{M=0}^{\infty }\sum_{\eta =-,+}\left[ \sqrt{M+1}\vert \Lambda _{\eta ,M+1}^{2}\rangle \langle \Lambda _{\eta ,M}^{2}\vert \right.   \notag \\
&&\left. -\beta _{2}\vert \Lambda _{\eta ,M}^{2}\rangle \langle \Lambda _{-\eta ,M}^{2}\vert +\text{H.c.}\right].
\label{H_DSC}
\end{eqnarray}

If we set $\omega _{d}=\omega _{1}+\omega _{2}$, assuming $\Omega =\omega_{q}^{1}=\omega _{q}^{2}$ and $\beta =\beta _{1}=\beta _{2}$, we can neglect fast-oscillating terms of Eq.~(\ref{H_DSC}) via RWA, as long as $\alpha_{0}/\omega _{\mu }$, $\alpha _{0}\beta /\omega _{\mu }$, $\alpha _{0}\beta^{2}/\omega _{d}\ll 1$. In addition, if one of the subsystems is initially prepared in the first excited state, while the other one in the ground state (for instance $ \vert \Lambda _{-,0}^{1} \rangle \otimes  \vert \Lambda _{+,0}^{2} \rangle $), the dynamics can be effectively described by the following Hamiltonian in the interaction picture with respect to $H_{0}=H_{R}^{1}+H_{R}^{2}$,
\begin{eqnarray}
\mathcal{H}_{\text{DSC}}^{I\text{(eff)}} &\approx &\frac{\alpha _{0}}{2} \sum_{N=0}^{\infty } ( N+1 )  \vert \Lambda _{-,N+1}^{1} \rangle  \langle \Lambda _{-,N}^{1} \vert   \notag
\\
&&\otimes  \vert \Lambda _{+,N+1}^{2} \rangle  \langle \Lambda_{+,N}^{2} \vert +\text{H.c.}  \label{H_DSC_eff}
\end{eqnarray}

The Hamiltonian above shows us that the system dynamics is effectively restricted to a Hilbert subspace constituted by a harmonic ladder $ \{  \vert \Lambda _{-,N}^{1} \rangle \otimes  \vert \Lambda
_{+,N}^{2} \rangle  \} $ with fundamental frequency $\varpi =\omega _{1}+\omega _{2}$, similarly to the SC case, but the interaction term of such effective Hamiltonian in this case satisfies a SU($1$,$1$)
symmetry \cite{Puri}, leading the whole system to a two-mode squeezed-like state with a time-dependent squeezing parameter $\xi  ( t ) =i\alpha _{0}t/2$. In other words, the hybrid subsystems effectively behave as bosonic modes entangled via DCE. 

\begin{figure}[t]
\includegraphics[width=0.45\textwidth]{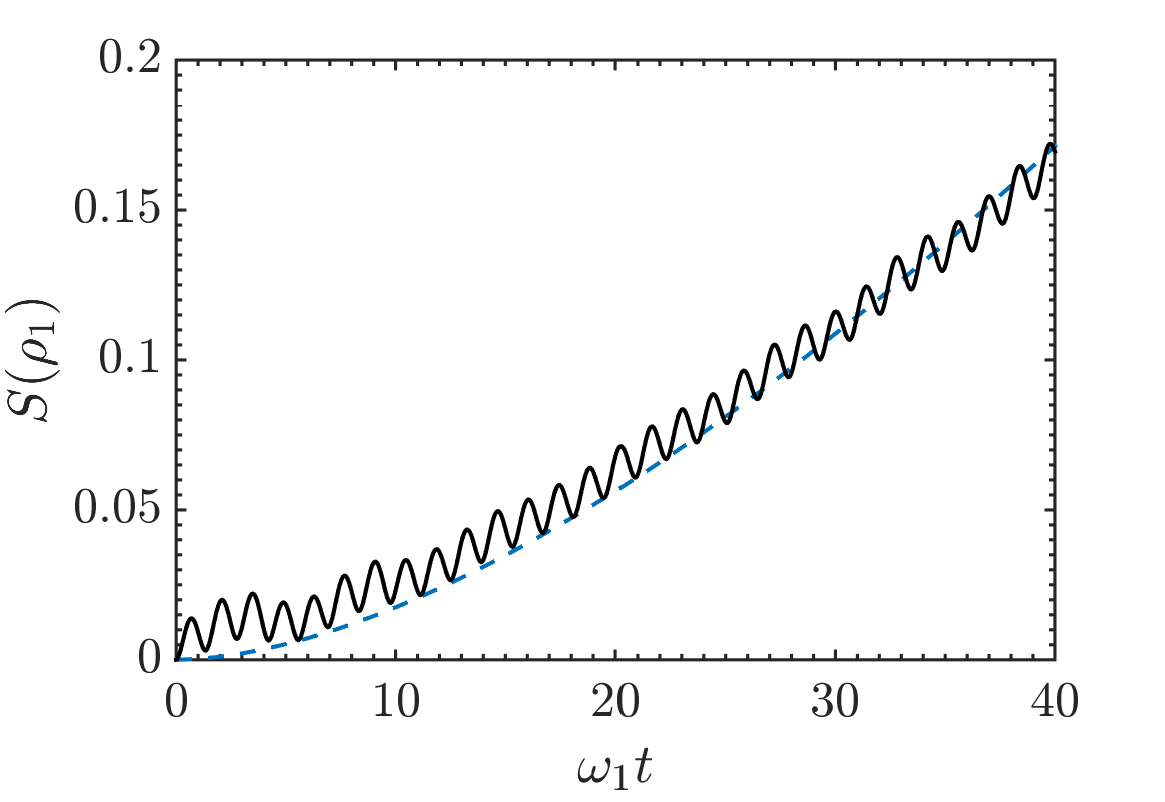}
\includegraphics[width=0.45\textwidth]{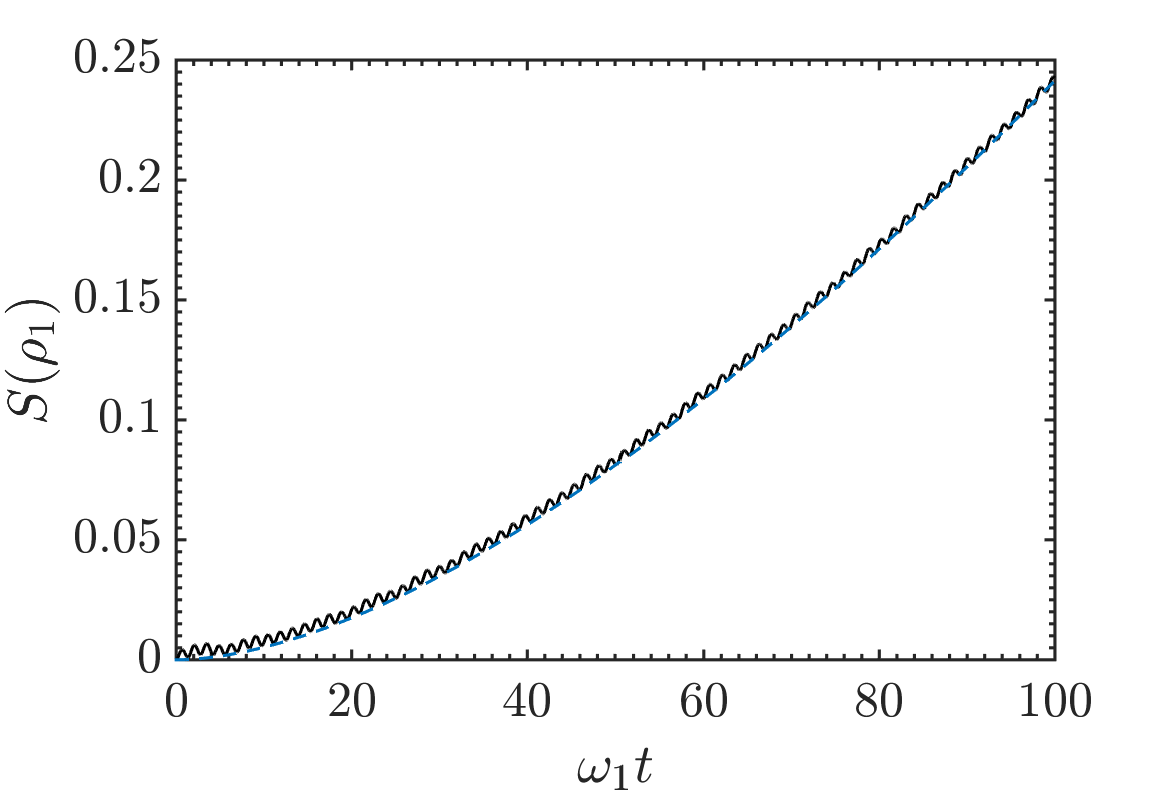}
\caption{Von Neumann entropy of the first hybrid subsystem calculated via Eq. (\ref{H_total}) (solid line) and Eq. (\ref{entropy}) (dashed line) using $\omega _{2}=1.25\omega _{1}$, $\Omega =\omega _{1}$, $\beta =1.5$ and $\alpha _{0}=0.01\omega _{1}$ (upper panel), and $ \omega _{2}=1.25\omega _{1}$, $\Omega =0.1\omega _{1}$, $\beta =1.5$ and $\alpha _{0}=0.005\omega _{1}$ (lower panel).}
\label{entvnnn}
\end{figure}

As the whole system is in a pure state, the entanglement between the hybrid systems can be measured through the von Neumann entropy of one of them, $S( \rho _{\mu }) $, where $\rho _{\mu }$ is the reduced density operator of the hybrid system $\mu $. Then, we have, for a two-mode squeezed state~\cite{Knight},

\begin{equation}
S( \rho _{1}) =-\sum_{N=0}^{\infty }\left[ \frac{\tanh^{2N}\vert \xi \vert }{\cosh ^{2}\vert \xi \vert } \right] \ln \left[ \frac{\tanh ^{2N}\vert \xi \vert }{\cosh^{2}\vert \xi \vert }\right] =S( \rho _{2}) .
\label{entropy}
\end{equation}
In Fig. \ref{entvnnn}, we show the von Neumann entropy of one of the hybrid system obtained via both Eq. (\ref{H_total}) and Eq. (\ref{entropy}), using $ \omega _{2}=1.25\omega _{1}$, $\Omega =\omega _{1}$, $\beta =1.5$ and $\alpha _{0}=0.01\omega _{1}$ (upper panel). The good agreement between the two entropies validates the effective description of the system dynamics by Eq. (\ref{H_DSC_eff}) at least until $\omega_{1}t = 40$. After this time, the SQUID starts to drive the system to excited states which are not well described by the considered approximation. On the other hand, when we deal with parameters which fulfill better the conditions for applying the approximation, our effective description can hold for longer times, as depicted in the lower panel of Fig. \ref{entvnnn}, where we consider $ \omega _{2}=1.25\omega _{1}$, $\Omega =0.1\omega _{1}$, $\beta =1.5$ and $\alpha _{0}=0.005\omega _{1}$.

Analogously to the case of SC regime, dissipative effects do not preserve the system dynamics in the harmonic subspace, destroying pairs of correlated excitations generated via DCE. Therefore, such
decoherence gradually decreases the entanglement between the polaritonic subsystems.

\section{Conclusions}

We have reported a theoretical proposal for entanglement generation between two qubit-cavity systems via DCE for the entire qubit-resonator coupling range. First, we have shown a semi-analytic explanation for the entanglement generation between the superconducting qubits when they are strongly coupled to their respective cavities. Then, we have extended such ideas for both USC and DSC regimes. Our results show the generation of a polaritonic Bell state in the USC regime, since each dressed qubit-resonator subsystem behaves as an effective two-level system mutually interacting through a DCE induced $XY$ coupling. On the other hand, in the DSC regime, the system can reach a polaritonic two-mode squeezed-like state, as the subsystems can behave as bosonic modes. Moreover, we have also shown the robustness of the entanglement generation against dissipative effects. In this manner, we have theoretically proved the feasibility of generating highly entangled states with the state-of-the-art technology in superconducting circuits for the SC and USC regimes, while foreseeing similar properties for the DSC regime.

\begin{acknowledgments}
We acknowledge support from the S\~{a}o Paulo Research Foundation (FAPESP) Grant No. 2014/24576-1, Spanish MINECO grant FIS2012-36673-C03-02, Basque Government grant IT472-10, UPV/EHU grant UFI 11/55, PROMISCE and SCALEQIT EU projects. E. S. also acknowledges support from a TUM August-Wilhelm Scheer Visiting Professorship and hospitality of Walther-Mei{\ss}ner-Institute and TUM Institute for Advanced Study.
\end{acknowledgments}

\end{document}